# Extraordinarily large intrinsic magnetodielectric coupling of Tb member within the Haldane spin-chain family, $R_2BaNiO_5$

Sanjay Kumar Upadhyay, P.L. Paulose and E.V. Sampathkumaran*

*Tata Institute of Fundamental Research, Homi Bhabha Road, Colaba, Mumbai 400005, India*

Abstract

The Haldane spin-chain compound, $Tb_2BaNiO_5$, has been known to order antiferromagnetically below ($T_N$=) 63 K. The present magnetic studies on the polycrystals bring out that there is another magnetic transition at a lower temperature ($T_2$=) 25 K, with a pronounced magnetic-field induced metamagnetic and metaelectric behavior. Multiferroic features are found below $T_2$ only, and not at $T_N$. The most intriguing observation is that the observed change of dielectric constant ($\Delta\varepsilon'$) is intrinsic and largest (e.g., ~ 18% at 15 K) *within this* Haldane spin-chain family, $R_2BaNiO_5$. Taking into account the fact that this trend (that is, the largest value of $\Delta\varepsilon'$ for Tb case within this family) correlates well with a similar trend in $T_N$ (with the values of $T_N$ being ~55, 58, 53 and 32 K for Gd, Dy, Ho and Er cases), we believe that the explanation usually offered for this $T_N$ behavior in rare-earth systems is applicable for this $\Delta\varepsilon'$ behavior as well. That is, single-ion anisotropy following crystal-field splitting is responsible for the extraordinary magnetodielectric effect in this Tb case. To our knowledge, such an observation was not made in the past literature of multiferroics.

## I. INTRODUCTION

The area of research exploring the coupling between magnetic and electric degrees of freedom in magnetic insulators continues to be at the center stage in condensed matter physics, ever since spin-induced ferroelectricity was reported in TbMnO$_3$ by Kimura et al [1], as magnetism and ferroelectricity were historically considered to be mutually exclusive. Besides, this cross-coupling bears relevance to applications, e.g., for novel data storage devices for writing electrically and reading magnetically. Various concepts have been proposed to explain the 'multiferroic' behavior arising from this cross-coupling, mainly in terms of asymmetric Dzyaloshinskii–Moriya interaction, spin-dependent *p-d* hybridization, and exchange-striction (see a review article by Dong et al [2]). Despite such an advancement of knowledge, there is a realization for the exploration of new mechanisms, territories [2] and pathways leading to enhanced coupling between magnetic and electric dipoles [also see, Refs. 3 and 4]. In this respect, recent investigations on thin films, heterostructures and surfaces of certain dilute magnetic semiconductors bring out the role played by magnetocrystalline anisotropy due to preferential orientation of magnetization on the cross-coupling, as reviewed in Ref. 2. It is worthwhile to bring out such a role of anisotropy among bulk form of compounds. It is also necessary to make sure that interference from issues due to other extrinsic contributions like leakage current is absent in such materials while interpreting the data, to enable reliable conclusions.

The above scenario prompted us to address a question how single-ion magnetic anisotropy of a given rare-earth (R) ion influences the cross-coupling effects. A comparison of this coupling within a given rare-earth series provides an opportunity to address this question, as a change of $R^{3+}$ does not perturb the lattice, barring lanthanide contraction. In this respect, the insulating Haldane spin-chain family R$_2$BaNiO$_5$ (Ref. 5) is ideally suited. These compounds crystallizing in *Immm*-type orthorhombic structure have been of special interest in the field of magnetism for over a quarter century for anomalies associated with Haldane spin-chain gap and magnetic ordering of Ni and (magnetic-moment containing) R ions at the same temperature (see, for instance, Refs. 5-24). Though this family has been considered to be prototype for Haldane spin-chain behavior, there was very little focus to understand magnetoelectric behavior of these insulators till recently. It may be noted that very high dielectric permittivity at room temperature for non-magnetic Y$_2$BaNiO$_5$ [17] and linear magnetoelectric coupling for magnetically ordering Ho analogue [18] were reported. This family of materials is characterized [6, 7] by infra-red allowed optical phonon mode with low frequency, i.e., the existence of a soft phonon mode which is a clue [25] for exploring magnetoelectric coupling. In light of this, in recent years, we have subjected many members of this family to intense studies to explore the cross-coupling phenomenon and reported [7, 19 – 23] a variety of magnetoelectric anomalies. While these compounds are found to exhibit multiferroic anomalies and magnetodielectric coupling, the point we would like to stress is that the observed magnitude of magnetodielectric coupling is less than a few percent (<< 4 %), even at fields as high as 140 kOe in all these compounds investigated till now. However, there was no attempt previously to investigate the Tb member, for which the Néel temperature ($T_N = T_1 = $ ~ 63 K) is the highest within this family. Such a high value within a given R series has been known [26-28] to be a consequence of single-ion anisotropy on magnetism. Naturally, this compound is suited to the aim outlined above.

We have therefore carried out exhaustive magnetic and magnetoelectric studies as a function of temperature (*T*) and magnetic-field (*H*) on the title compound. It may be stressed that, barring initial magnetic susceptibility ($\chi$) and neutron diffraction measurements [11, 24], reporting onset of antiferromagnetic order, there is no further report in the literature on this compound. The present results reveal the existence of another magnetic transition below ($T_2=$) 25 K, with magnetic-field induced spin-flop effects in isothermal magnetization (*M*) and magnetodielectric coupling as well as multiferroic behavior. The key observation being stressed is that this compound exhibits the largest intrinsic magnetodielectric effect *within this family* below $T_2$, e.g., ~18% at 15 K. We therefore suggest that single-ion (4f-orbital) anisotropy plays a role on the magnetodielectric properties in this family.

## II. EXPERIMENTAL DETAILS

The polycrystalline specimen of $Tb_2BaNiO_5$ was prepared by a standard solid-state reaction route as described in Ref. 24. Stoichiometric amounts of NiO (99.995%), $BaCO_3$ (99.997%) and $Tb_2(CO_3)_2 \cdot nH_2O$ (99.9%) were used as initial precursors. All these starting materials before weighing were dried for 2 hours to remove traces of moisture. The pellet of the mixture of these dried starting materials was first heated at 950 $^0$C for 12 hours, followed by sintering at 1050 $^0$C, 1150 $^0$C, 1250 $^0$C and 1350 $^0$C for 12 hours each with intermittent grindings. These sinterings were done in the flow of high-purity Ar, because of instability of $Tb^{3+}$ at high temperatures in air. The formation of the sample was confirmed by x-ray powder diffraction pattern using Cu-$K_\alpha$ radiation at room temperature. The observed diffraction pattern was refined by Rietveld fitting by using Fullproof program. The lattice parameters ($a$= 3.781(3) Å, $b$= 5.799 (2) Å and $c$= 11.411 (4) Å) match well with the literature [24]. The homogeneity of the sample was further confirmed by scanning electron microscope and the composition (within 2%) was confirmed by energy-dispersive x-ray analysis. Instruments used for measurements of magnetization, heat-capacity (*C*), and dielectric permittivity as a function of *T* down to 1.8 K and in magnetic-fields have been described in our earlier publications [7, 19-23]. Bias electric-field (*E*) measurements were done with the help of a Keithley 6517B electrometer. In this case, the sample was cooled to 5 K in the absence of electric-field and the bias current, $I_B$, was measured in the presence of a bias electric field of 2 kV/cm with a heating rate of 2 K/min. Isothermal *H*-dependence of dielectric constant ($\varepsilon'$) was measured at some selected temperatures.

## III. RESULTS AND DISCUSSION
### A. Evidence for antiferromagnetic transitions at 63 and 25 K

Our dc magnetic susceptibility ($\chi$) data obtained as a function of *T* are in good agreement with Ref. 24. There is a kink near 64 K due to the onset of magnetic ordering concurrently from Tb and Ni sublattices [11]. In addition, there is a broad peak at ~ 40 K (figure 1a), which is usually attributed to the persistence of 1-dimensional magnetic feature, characteristic of this family of Haldane spin-chain systems. Despite the weakness of the feature at $T_N$ in $\chi(T)$, heat-capacity reveals a prominent anomaly. A distinct jump in the plot of *C(T)* at 63 K (figure 1b) could be observed. The fact that the magnetic ordering is of an antiferromagnetic-type is confirmed by the observation that the peak gets gradually suppressed, shifting towards low-temperature range with increasing dc magnetic-field (figure 1b, inset). It is worth noting that the observed value of $T_N$ is the highest in this series; for the Gd, Dy, Ho and Er cases, respective $T_N$ value is equal to ~55, 58, 53 and 32 K [22, 19, 18, 7]. It was theoretically established long ago [26-28] that the anisotropy of the 4f orbital of the crystal-field split ground state (that is, the sign of some of the crystal-field terms to Hamiltonian) in general plays a role in such an enhancement of $T_N$ for a heavy member in a given rare-earth family.

In figures 1a and 1c, one can see dc χ obtained in a field of 100 Oe for zero-field-cooled (ZFC) and field-cooled (FC) conditions of the specimen and dχ/dT for the dc χ data obtained in a field of 5 kOe respectively. There is a distinct increase in dχ/dT at 25 K, as though there is a change in the magnetic character around this temperature. There is no notable bifurcation of ZFC-FC curves near (or below) 25 K, that could characterize this transition as spin-glass freezing. In fact, heat-capacity exhibits a well-defined peak at this temperature (with a jump as much as 10 J/mol K) (figure 1b), which renders a strong support to the existence of a non-glassy transition at this temperature. The fact that this $C(T)$ feature also shifts towards lower temperatures with increasing dc magnetic-fields (figure 2b) is consistent with antiferromagnetism.

Many members of this rare-earth family have been shown [7, 20-23] to reveal spin-glass characteristics at low temperatures (<10 K) well below respective $T_N$. In order to confirm the absence of spin-glass freezing in this case, we performed isothermal remnant magnetization ($M_{IRM}$) studies at 2, 5 and 20 K. That is, after cooling the specimen in zero-field to the desired temperature, a field of 5 kOe was switched on for 5 mins; then $M_{IRM}$ was measured as a function of time after switching off the field. $M_{IRM}$ was found to be negligible immediately after switching off the field and the slow decay behavior of $M_{IRM}$ expected for spin-glasses was found to be absent (not shown here). We have also measured ac χ with various frequencies (ν= 1.3, 13, 133 and 1333 Hz) and all these curves, resembling the dc χ(T) curve, overlap without any ν-dependence of the peak (not shown here). In addition, there is no evidence for the any feature in the imaginary part, which further establishes absence of spin-glass freezing.

In order to support the existence of a magnetic transition around 25 K, we measured isothermal magnetization at various temperatures. We observe (figure 3a) a distinct upturn in $M(H)$ near ($H_{cr}$=) 60 kOe with a very weak hysteresis for $T \le 15$ K, attributable to the existence of a spin-flop transition. Such a spin-flop transition was not known in the past literature for this compound. However, for 30 and 50 K, this field-induced transition is absent, thereby confirming that there is a difference in the magnetic character as the sample is warmed from 15 K to 30 K. At 75 K, no worthwhile feature is seen, as expected. The observed field-induced transition seems to be of a disorder-broadened first-order-type transition, as the virgin-curve lies outside the envelope curve (see the first quadrant of the hysteresis loop for 2 K shown in the inset of figure 3a).

In short, this is for the first time that such a second magnetic transition (around 25 K) is reported for this compound. The temperature at which such a second transition appears is the highest for Tb member, as the case for $T_N$. Garcia Matres [11] claimed on the basis of neutron diffraction data that there is a small moment (1.4 μ$_B$) on Ni (in addition to that on Tb), making an angle with c-axis at $T_N$; this Ni moment was proposed to rotate further along c-axis with decreasing temperature; Tb moment was proposed to align along c-axis below $T_N$. In view of our present finding, it is worthwhile to reinvestigate this compound across 25 K carefully by neutron diffraction.

### B. Dielectric and magnetodielectric behavior

In figure 2b, we show ε′ (T), obtained with an ac bias of 1 V and ν= 100 kHz in various magnetic-fields. It is clear that there is no worthwhile feature at $T_N$. The zero-field and in-field curves look similar, almost overlapping with each other over a wide temperature above $T_2$. This means that magnetodielectric effect is negligible in the vicinity of $T_N$. This was verified by isothermal magnetodielectric data as well (see below). However, as the temperature is lowered below $T_N$, a distinct peak appears at $T_2$. Besides, with increasing $H$, the peak shifts gradually to a lower temperature, for instance, to ~ 24, 22, and 19 K for $H$= 40, 80 and 140 kOe respectively.

All these findings establish the existence of magnetodielectric coupling. This *H*-dependence of the peak in ε′ resembles that seen for the peak in *C(T)* (figure 2a). We could not resolve any frequency dispersion of the peak in ε′ for all *H*, thereby ruling out any kind of glassy behavior (see the overlapping curves for 1 and 100 kHz in the inset of figure 2b). This finding is in contrast to glassy electric dipole behavior (that is, frequency dispersion) seen for most members of this series [7, 20-23] well below $T_N$. The observed coupling must be intrinsic, as the value of the loss factor (tanδ) is insignificant (of the order of 0.0003) in the temperature range of interest.

Further support for magnetodielectric coupling is obtained from isothermal magnetodielectric data, shown in figure 3b. For 5K, Δε′ [defined as = {ε′(*H*) - ε′(0)}/ε′(0)] undergoes a weak increase with *H* initially, and close to $H_{cr}$, there is a prominent upturn, supporting the existence of a (broad) meta-electric transition [29]. The field at which this upturn happens decreases with increasing temperature; see, for instance, the curves for 5, 10, and 15 K. The curves are weakly hysteretic in a wider *H*-range around $H_{cr}$. Besides, the virgin curve lies outside the envelope curve, as demonstrated for 2 K (figure 3b, inset). No such metaelectric-like feature is observed for the curves well above 25 K (see the curve for 50 K in figure 3b) and the values are also found to be <1%. This behavior clearly tracks the features observed in *M(H)* curves. This one-to-one correspondence between Δε′ and *M(H)* offers conclusive evidence for the existence of magnetodielectric coupling below $T_2$. A careful look, however, at the Δε′ curves in the range 2 – 10 K reveals that there is a plateau around 70 kOe followed another upturn. This signals the existence of another magnetic-field-induced transition, which is somehow smeared in *M(H)* curves. We therefore think that there is more than one magnetic-field transition in this Tb material and it is a common feature in this family of materials [19].

We now focus on the most important observation in magnetodielectric effect. That is, Δε′ (below $T_2$) attains larger values compared to that in other members of this series. For instance, Δε′ attains a maximum value of about 18% (for about 120 kOe). This is a significant finding for polycrystalline material, as extrinsic contributions do not play at low temperatures in this compound. This value is more than (or comparable to) that reported for many well-known polycrystalline magnetoelectric oxides, e.g., 7% for $EuTiO_3$ [Ref. 25], 10% for $TbMnO_3$ [Ref. 1], 13% at 10 K for hexaferrite $Ba_{0.5}Sr_{1.5}Zn_2Fe_{12}O_{22}$ [Ref. 30], and 16% for $CaBaCo_4O_7$ [Ref. 31]. In sharp contrast to this behavior of Tb compound, we find that the value does not even exceed a few percent for other R members, e.g., Nd, Gd, Dy and Er members [7, 19-23]. The fact that the Tb member alone stands out in this respect rules out any direct role of the Haldane spin-chain. It is possible that exchange-striction mechanism and/or magnetic-field-induced spin-current (or polarization) due to a change in the angle of the magnetic-moment with *c*-axis [11] is responsible for this large value [32]. Since the magnitude of Δε′ is not that large in other members of this series with similar magnetic structure [11] with Ni moment making an angle with c-axis, we are tempted to conclude that the anisotropy of the 4f orbital of Tb following crystal-field splitting plays a crucial role in enhancing magnetodielectric coupling, *just as it maximizes $T_N$ for this member in this family.* The fact that this enhanced coupling is favoured by the magnetic structure below $T_2$, and not by the one between $T_1$ and $T_2$, implies that the magnetic structure also should be favorable for the cross-coupling – a fact established in various models for spin-induced multiferroicity [2]. At this juncture, we would like to state that our preliminary x-ray diffraction experiments at selected temperatures establish the absence of any crystal structure change across 25 K and therefore the anomalies discussed above are magnetic in origin. Incidentally, in the case of $RMn_2O_5$ family as well, similar correlation between $T_N$ and magnetodielectric coupling exists, with a maximum for the Dy compound [33, 34].

The exact microscopic origin of the role of single-ion anisotropy to enhance cross-coupling features is not clear to us at present. We however like to state that Jia eta al [35] proposed a theory more than a decade ago how bond electric polarization develops due to non-collinear spin configuration in the limit of strong Hund coupling. Extrapolating this idea, we speculate that the crystal-field-split orbitals following magnetic ordering at low temperatures in these rare-earth systems facilitate the distortion of electron cloud surrounding the rare-earth ion, resulting in an electronic charge dipole. Such distortions may promote dramatically the exchange-striction effect discussed by us in Ref. 7.

### C. Electric polarization behavior

Since there is a well-defined peak in $\varepsilon'(T)$ at $T_2$, we looked for electric polarization anomalies as well at this temperature. The space group *Immm* is centrosymmetric and hence unfavorable for ferroelectricity. However, a small displacement of the atoms around $NiO_6$ octahedra, following exchange-striction, can result [7] in non-centrosymmetric space group, *Imm2*. We therefore searched for ferroelectricity. We followed the protocol suggested in the recent literature [36, 37] to look for intrinsic ferroelectric features, as stated in the experimental section. The magnitude of $I_B$ exhibits (figure 4a) a broad peak around 22 K. The variation of the relative value ($\Delta P$) of electric polarization derived from the $I_B$ data is plotted in figure 4b. The sign of polarization gets reversed with a change in the polarity of $E$. In addition, there is a pronounced shift of the onset of temperature of with increasing dc $H$ (see, for instance, the polarization curve for 140 kOe in figure 4b). This trend mimics $C(T)$ and $\varepsilon'(T)$ behavior (figures 2a and 2b). We are not able to resolve any shift in the peak temperature or onset temperature for different rates of warming (figure 4, inset), which is consistent with the absence of any other contribution like 'thermally stimulated depolarization' [38]. These findings suggest onset of ferroelectricity following the magnetic transition at 25 K (that is, multiferroic behavior below $T_2$). Possibly, such a transition also can contribute to the $C(T)$ jump.

### IV. SUMMARY

We have investigated the polycrystalline Tb-based compound $Tb_2BaNiO_5$ by magnetization, heat-capacity, dielectric permittivity and pyrocurrent measurements as a function of temperature and magnetic field. We find that, in addition to long-range antiferromagnetic ordering below 63 K, there is another magnetic ordering at 25 K. The magnetic transition is not of a glassy type – a situation different from that observed in other heavy rare-earth members. Another behavior that makes it different from other heavy rare-earth members is that this compound exhibits magnetoelectric coupling with multiferroic features below 25 K , and not at or above $T_N$. The key observation being stressed is that the observed magnitude (>18%) of magnetodielectric coupling is the largest for this Tb member within this rare-earth family, mimicking the pattern of Néel temperature. This observation offers a clue for the role of the single-ion anisotropy of the Tb ion in promoting magnetodielectric coupling for Tb member in this family.

Figure 1:
For $Tb_2BaNiO_5$, (a) dc magnetic susceptibility ($\chi$) obtained in 100 Oe for the zero-field-cooled (points) and field-cooled (red line) conditions. (b) Heat-capacity below 80 K. Inset shows how the peak in the vicinity of $T_N$ shifts with increasing magnetic-fields [shown for 0 (points), 20 (red line through points), 40, and 60 kOe). (c) The derivative of $\chi$, obtained in 5 kOe. Vertical arrows mark transition temperatures.

Figure 2:
(a) Heat-capacity behavior and (b) dielectric constant ($\nu$= 100 kHz), below 40 K. In the inset of (b), the data for 1 kHz (line) and 100 kHz (points) in the wider range till 80 K are shown to highlight the absence of $\nu$-dependence of the peak and of any feature in the dielectric constant at $T_N$. In the mainframe, the lines through the data points serve as guides to the eyes; the arrows are drawn to show the direction in which the curves move.

Figure 3:
(a) Isothermal magnetization per formula unit (f.u.) and (b) magnetodielectric behavior. The hysteresis loops at 2 K are shown in respective insets, to highlight that the virgin curve (red) lies outside the envelope curve and, for the sake clarity, the plot is restricted to the first quadrant in the $M(H)$ plot. The lines through the data points serve as guides to the eyes and the arrows are drawn to show the direction in which the magnetic-field was changed.

Figure 4:
(a) Variation of dc-biased current ($I_B$) with temperature (<40 K) obtained with a bias electric field of -2 kV/cm. (b) The change in the electric polarization derived from the $I_B$ for opposite polarities of electric-field in the absence of external magnetic-field; the polarization behavior in the presence of a dc magnetic field of 140 kOe, is also included. An arrow is drawn in (a) to show the direction in which the curves move with increasing fields. Inset shows the $I_B$ curves for different rates of warming.

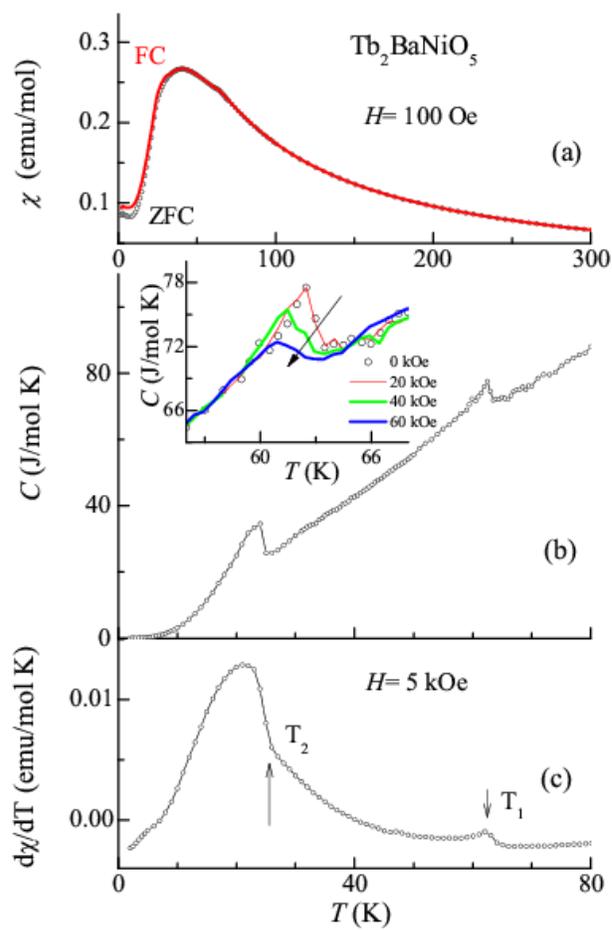

Figure 1

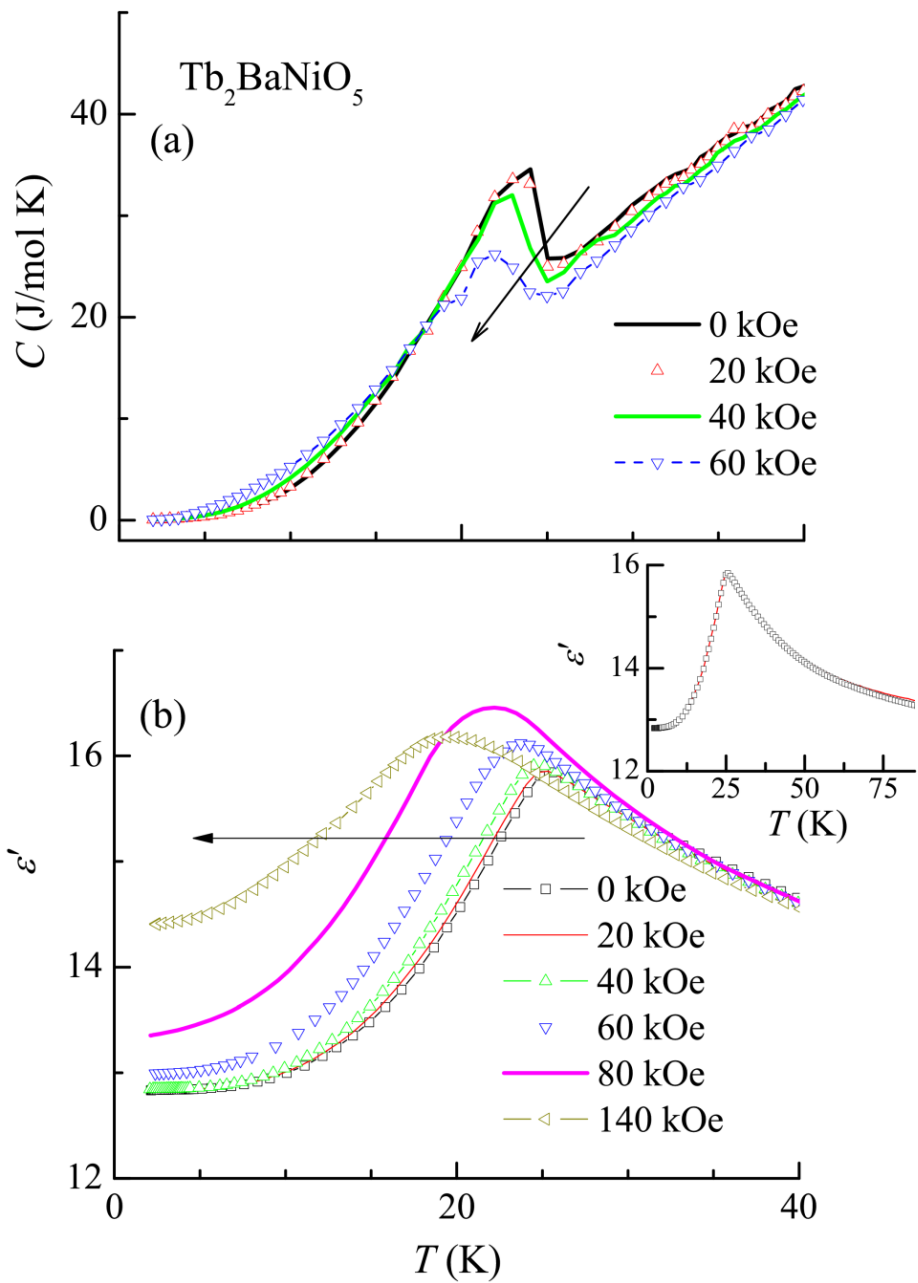

Figure 2

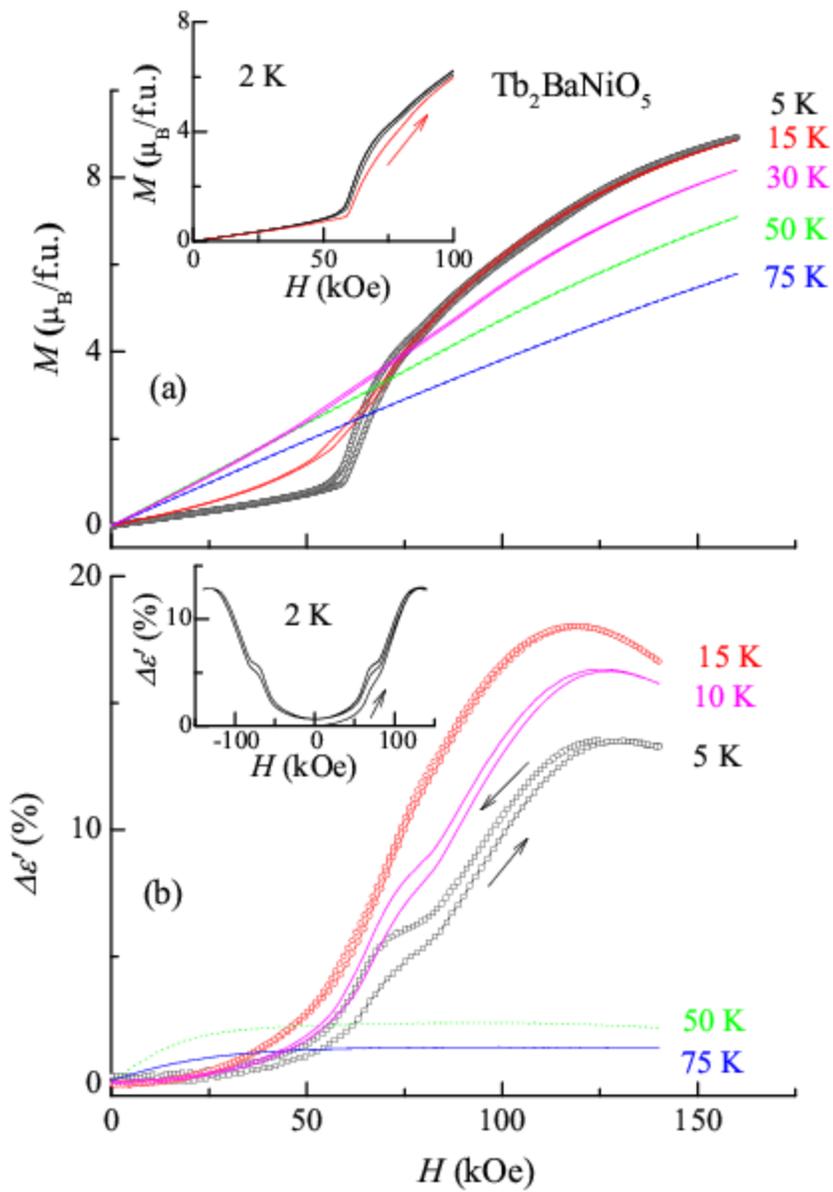

Figure 3

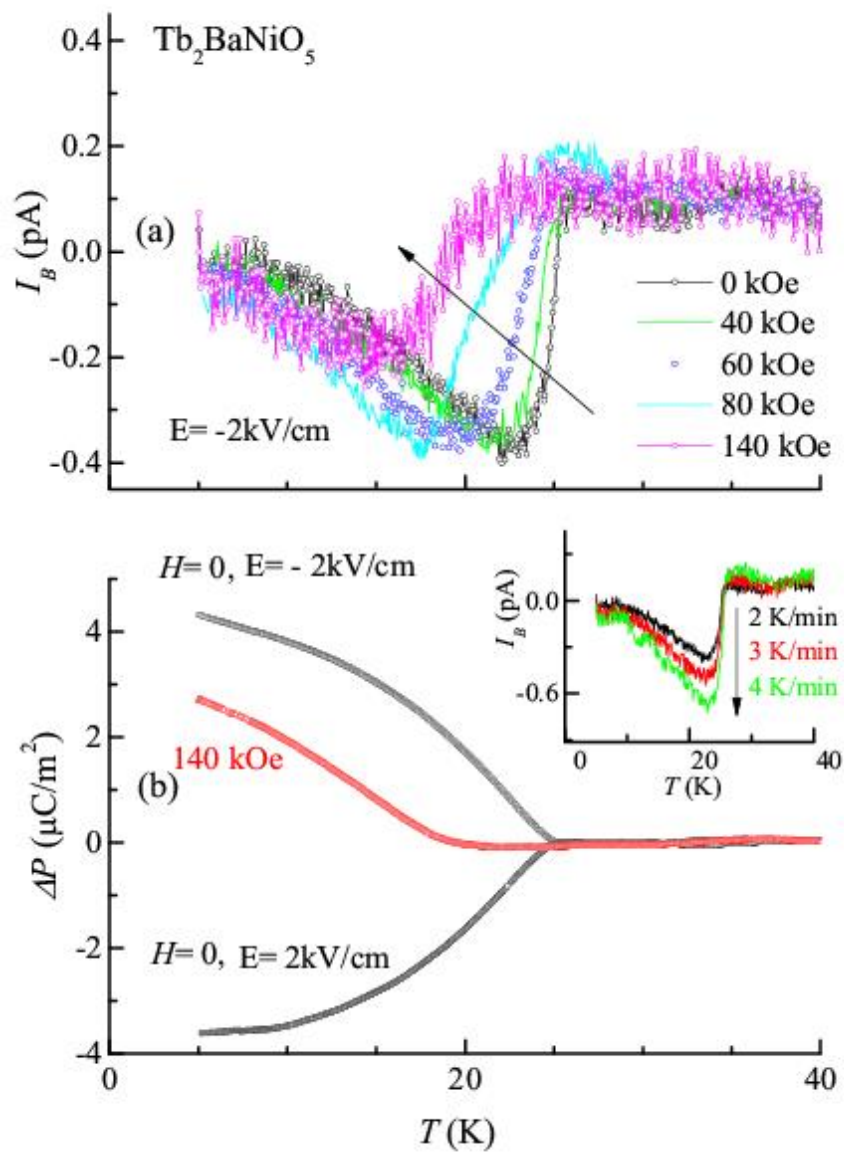

Figure 4